# Control of Magnetic Dipole Emission with Surface Plasmon Polaritons

**S. Mashhadi[1], M. Durach[2], D. Keene[1], N. Noginova[1]**

[1]*Norfolk State University, Norfolk VA 23504*
[2]*Georgia Southern University, Statesboro, GA 30460*
s.mashhadi@spartans.nsu.edu

**Abstract:** The possibility to use surface plasmon polaritons for enhancement of weak magnetic dipole transitions is analyzed theoretically and demonstrated experimentally for simple flat geometry and sine-wave profile modulated plasmonic films. Spontaneous emission of $Eu^{3+}$ in organic matrices deposited onto plasmonic surfaces demonstrates specific angular and polarization patterns at both electric and magnetic dipole transitions with a well-defined maximum at the plasmon decoupling conditions manifesting the character and magnitude of the effect.

**1. Introduction**

Plasmonic systems and metamaterials attract much attention due to novel optical effects predicted and observed in these systems, which provide numerous opportunities for fundamental and applied research, including new approaches for light manipulation [1-3], control of quantum processes via modified local environments [4-6] and many promising applications in technology [7,8]. Significant enhancement in spontaneous emission of electric dipoles in plasmonic environment [9-13] is one of the most studied effects; it is commonly discussed in terms of the Purcell factor [14] and associated with strong optical electric fields in the vicinity of plasmonic structures.

Another important property associated with nanostructured plasmonic systems is optical magnetism, i.e. a strong modification of optical magnetic fields in or in close vicinity to structures [15-18]. Local magnetic fields can be probed via a "magnetic Purcell effect" [18] observing changes in magnetic dipole emission intensity, which is modified in enhanced optical magnetic fields analogous to the Purcell factor-related enhancement of electric dipole transitions. As was discussed in [19] magnetic dipole and electric dipole emitters can serve as spectroscopic probes for characterization of plasmonic structures and metasurfaces, providing a direct opportunity to experimentally probe and map optical electric and magnetic fields.

Spontaneous emission of magnetic dipoles in modified optical environments was the subject of numerous theoretical and experimental studies [19-29]. Most studies consider flat geometry and relatively large distances between emitters and the metal, where the effects are primarily associated with the modification of far-field optical modes. However, more studies are needed to better understand the effects of magnetic dipole emission in very close vicinity to plasmonic metals where coupling with plasmonic modes plays a defining role for both electric and magnetic emitters.

Enhancement of magnetic dipole emission by resonant nanostructures such as gold nanoholes has been demonstrated as well [30, 31] but is still largely unexplored. Recently, it was shown that dielectric nanostructures such as silicon spheres [32], squares [33] or disks [34, 35] are very promising for magnetic dipole enhancement. While in the infrared range the enhancement can be significant [35], in the optical range the Q-factors of such resonant structures are usually low, which results in a relatively small enhancement. On the other hand, propagating surface plasmon polaritons (SPP) in flat films have higher Q-factors and can provide significant effects on the emission of dipoles located in their vicinity. In our work, we theoretically analyze how the effects depend on the type of the emitters (electric or magnetic) and distance between emitter and metal, and experimentally demonstrate the SPP-related emission enhancement in a simple flat geometry and sine-wave profile modulated metal films.

Optical transitions associated with magnetic dipoles (MD) can be found in the spontaneous emission of quantum dots, some transition ions, and rare earth ions [36-38]. They are commonly weak, and the possibility to significantly enhance them could open applications in displays, bioprobes, and quantum light sources. Trivalent europium is one the most common materials for MD transition studies and applications since the 1940s when europium ions ($Eu^{3+}$) were used in wide-angle interference experiments for the first time [39]. Organic systems with $Eu^{3+}$ ions [40] are very convenient systems for studying magnetic enhancement. They are highly luminescent, can be pumped in the ultra-violet range, and have well-resolved distinct peaks in the spontaneous emission associated with primarily electric and primarily magnetic dipole transitions. Observing changes in the intensities and radiation patterns of the magnetic transition, one can estimate the effects of the local environment and any possible magnetic enhancement.

## 2. Theoretical modeling

For simplicity, we consider an idealized situation, where electric and magnetic dipoles in air are oriented perpendicularly or parallel to the flat gold film with 30 nm thickness, see Fig. 1 (a). Photonic fields, E and H as a function of the in-plane component of the optical k-vector, $k_x$, $(0 < k_x < \infty)$ are calculated using the dyadic Green's function approach [9] assuming a wavelength of 600 nm and the corresponding permittivity of gold [41]. The power emitted by electric and magnetic dipoles into photonic modes in the range $(k_x, k_x + dk_x)$, is estimated using

$$\frac{dP_e}{dk_x} = \frac{ck_0}{2} \cdot Re\left[\frac{k_x}{k_z}\left(\frac{p_\parallel^2}{2} \cdot \left[k_z^2(1 - R_p e^{2ik_z h}) + k_0^2(1 + R_s e^{2ik_z h})\right] + p_\perp^2 \cdot k_x^2(1 + R_p e^{2ik_z h})\right)\right],$$

$$\frac{dP_m}{dk_x} = \frac{ck_0}{2} \cdot Re\left[\frac{k_x}{k_z}\left(\frac{m_\parallel^2}{2} \cdot \left[k_z^2(1 - R_s e^{2ik_z h}) + k_0^2(1 + R_p e^{2ik_z h})\right] + m_\perp^2 \cdot k_x^2(1 + R_s e^{2ik_z h})\right)\right],$$

such that the total power emitted by the dipole $P_e = \int_0^\infty \left(\frac{dP_e}{dk_x}\right) dk_x$ is

$$P_e = \frac{ck_0}{2} \text{Im}(\boldsymbol{p}^* \cdot \hat{G}_e(\boldsymbol{r}_0, \boldsymbol{r}_0) \cdot \boldsymbol{p}), \quad P_m = \frac{ck_0}{2} \text{Im}(\boldsymbol{m}^* \cdot \hat{G}_m(\boldsymbol{r}_0, \boldsymbol{r}_0) \cdot \boldsymbol{m}).$$

We show the distribution of emitted power over photonic modes in Fig. 1(b). This distribution has a well-defined peak at $k_x$ = 1.013, manifesting the emission to the SPPs [Fig. 1(c)]. The total probabilities for the emission to far-field ($P_{FF}$), SPP ($P_{SPP}$) and high-k lossy modes ($P_{NF}$) are estimated with the integration over $0 < k_x < 1$, $1 < k_x < 1.03$, and $k_x > 1.03$ correspondingly, and normalized to the total emission $P = P_{FF} + P_{SPP} + P_{NF}$. We also show the modification of the total emitted power $P$ as compared to power of emission in vacuum $P_0$, which is equivalent to similarly normalized modification of the rate of emission $\gamma$.

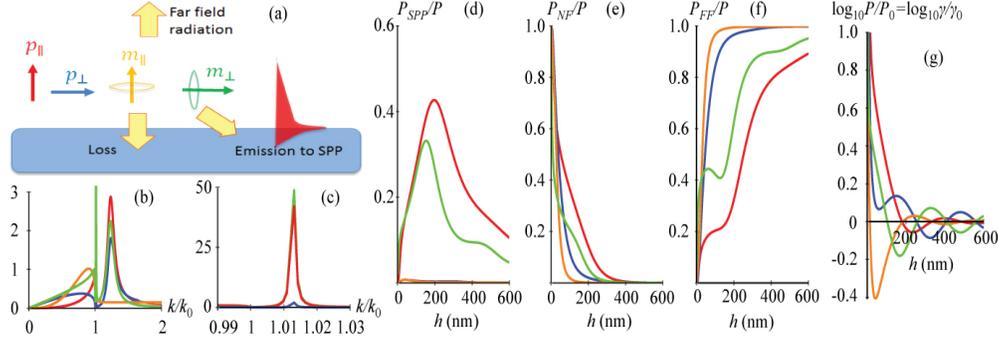

Fig. 1. (a) Schematics for simulations: dipole types and orientations, and pathways of energy flow; (b) example of the calculations results, $\frac{k_0}{P_0}\frac{dP_e}{dk_x}$ for electric vertical (red), electric parallel (blue), magnetic vertical (orange), magnetic parallel (green) as the function of $k_x$, at the distance 25 nm from a 30 nm gold film; (c) the same as (b), but with SPP peak zoomed on; (d-f) Probabilities of (d) excitation of SPP, (e) near field excitation of high-k modes, (f) far-field radiation for certain dipole type and orientation (indicated with corresponding color) as the function of the distance from the plasmonic film; (g) rate of emission for different types of dipoles normalized to that in vacuum in logarithmic scale.

As one can see, both types of dipoles can efficiently excite SPPs, with the probabilities highest for electric dipoles oriented perpendicular to the interface and magnetic dipoles parallel to the interface. The maximum for the magnetic excitation of SPP is shifted toward lower distances in comparison with the electric excitation. The excitation of high-k modes for magnetic emitters has a different distance dependence and is generally lower than that for electric emitters, indicating that smaller distances between emitters and a plasmonic metal can be preferable for experimental tests. Note that in the real experiment one needs to take into account the refractive index of the medium with emitters and the arbitrary orientation of dipoles.

## 3. Experimental

In order to probe the SPPs excitation by emitters in the vicinity of a plasmonic metal, the SPPs should be decoupled to far-field. In our experiments we use two different approaches. The schematics of the samples (Structures S1 and S2) are shown in Fig. 1 (a). In the first structure, emitters are placed in the vicinity of a gold surface with a periodically modulated height profile (gratings). If the modulation amplitude is relatively low, the SPP dispersion curve and its Q-factor are practically the same as in flat films [42]. The decoupling angle, $\theta$, is defined by the condition, $k_{ss} = k_0 \sin\theta + n\, 2\pi/d$, where $d$ is the period of the structure, $k_{spp}$ and $k_0$ are the SPP and optical k vectors correspondingly, and $n$ is an integer. As the second system, we use a flat silver film deposited onto a high-index prism, where the SPP decoupling angle should correspond to the SPPs excitation angle in Kretschmann geometry.

The profile-modulated gratings (S1) are fabricated with the holographic lithography technique [43]. The square substrates (2.5 cm) are immersed in 2% Hellmanex III solution followed by an ultrasonic bath for 20 min; the same procedure is repeated twice using DI water and Ethanol to provide a dust free substrate. Then substrates are spin coated with photoresist S 1805 diluted with propylene glycol monomethyl ether acetate (PGMEA), followed by soft-baking at 98° C for 1min. The photoresist layer is exposed to an interference pattern in a holographic set-up using a He-Cd laser beam ($\lambda = 325$ nm). Next, the samples are developed with MF-26A. Gold with thickness $l_g = 60$ nm is deposited on top of the gratings using the thermal evaporation method. Atomic Force Microscopy (AFM) is used for the characterization of the fabricated sample and determination of its parameters. The second sample (S2) is a high index prism covered by 30 nm-thick silver using thermal deposition.

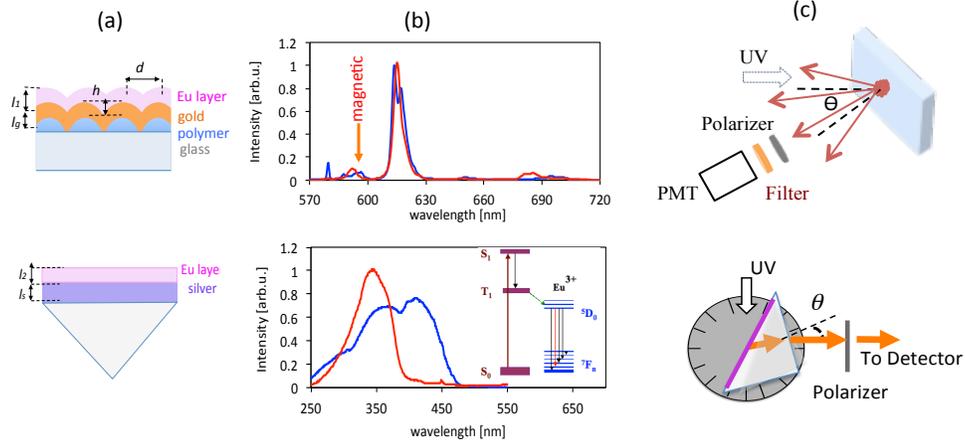

Fig 2. (a) Schematics of the experimental samples, S1 (top) and S2 (bottom). (b) Emission and excitation spectra of $Eu^{3+}$ in Eu $(NO_3)_3Bpy_2$ (red) and $Eu(TTA)_3(L1a)$ (blue). The inset shows schematics of the levels and transitions. (c) Setups for experiments with the profile modulated (top) and flat (bottom) systems.

As emitting layers, we test two different chromophore materials with $Eu^{3+}$, $Eu(NO_3)_3Bpy_2$ and $Eu(TTA)_3(L1a)$ synthesized in house following [19, 44] and [45] correspondingly. They both can be excited in the UV and display bright red emission in the visible range, with the strongest line at $\lambda \approx 611$ nm, corresponding to the electric dipole transition $^5D_0 \rightarrow {}^7F_2$. The line at $\lambda \approx 590$ nm is related to the magnetic dipole transition [40], $^5D_0 \rightarrow {}^7F_1$, Fig 1 (b).

$Eu(NO_3)_3 \cdot Bpy_2$ crystals and PVP(10 wt.%) are dissolved in a water/ethanol mixture and deposited onto the profile-modulated film using the spin coating method producing a layer with thickness $l_1 = 60$ nm. A thin layer of $Eu(TTA)_3(L1a)$ is deposited on the prism over the silver film using a modified Langmuir-Blodgett technique, which produces films with an essentially uniform thickness [21]. A 20 microliter of the solution $Eu(TTA)3(L1a)$ and polystyrene in chloroform is spread on the water's surface. After evaporation of the chloroform, a thin film is formed on the water's surface. By immersing the prism into the water, the film is transferred to the prism. The thickness of the film is measured by replicating the same film on a glass substrate and using a profilometer. It is found to be of approximately 20 nm thick.

The experimental setups are illustrated in Fig. 1 (c). In the first experiment, the emission is excited with the light from a He - Cd laser ($\lambda = 325$nm) which is brought to the gratings (S1) by an optical fiber. The emission in s and p polarizations is collected as the function of angle. In order to record the electric and magnetic emission separately, interferometric filters with transmission windows centered at 610 nm and 590 nm correspondingly are inserted into the recording channel. The emission from the flat film (S2) is studied in the fluorometer setup (Fluoromax-3). The sample is placed on rotational stage and excited by $\lambda = 330$ nm. The emission spectra are recorded in the 570 nm - 630 nm spectral range in s and p polarizations at multiple positions of the goniometer corresponding to the different observation angles $\theta$.

**Results and discussion**

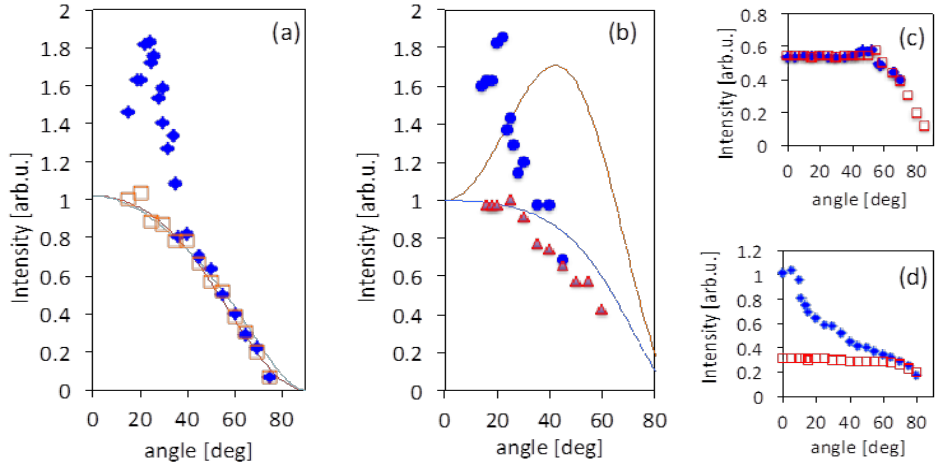

Fig.3. Emission from the gratings S1 at s (red) and p (blue) polarizations and various parameters of the structures. (a) Electric dipole transition (610 nm), $d$= 350 nm, $l_l$ = 60 nm; (b) magnetic dipole transition (590 nm), $d$= 350 nm, $l_l$ = 60 nm; (c) l = 610 nm, d= 350 nm, $l_l$ = 150 nm; and (d) l = 610 nm, $d$= 420 nm, $l_l$ = ~ 40 nm. Points are experiment, solid lines are calculations of far-field emission.

The results obtained in the profile modulated films are shown in Fig. 3. As one can see, both electric (Fig. 3(a)) and magnetic (Fig. 3(b)) emissions are quite similar to each other, and show a well-defined peak, which is observed only in p polarization at ~ 24° at the wavelength of the electric dipole transition and ~ 22° at the magnetic dipole transition. Such a strong difference in s and p polarizations is observed only if the Eu films are thin enough. In Figure 3 (c), in the experiment with a thick Eu layer, no prominent difference is observed between the two polarizations. A small maximum around 50° can be attributed to the effect of diffraction [46].

In order to prove that the peak observed in (a, b) corresponds to the SPP decoupling conditions, we test the grating structure with a different modulation period, and observe a similar peak at smaller angles (~ 5°). In Figs. 3(a, b), we also try to compare the experimental data with the theoretical predictions for the far-field emissions from a flat film, calculated with averaging over the thickness of the polymer film and dipole orientations. Experimental data for the electric dipole emission can be well fitted with the theory except for the additional peak since the SPPs are not decoupled from the flat film. Predictions for the magnetic dipole emission do not well correspond to the experiment (which may be due to the fact that the much stronger electric transitions occur from the same level which is not taken into account in this model), however they show that the additional peak observed at p polarization could not be explained with only the modification of far-field modes near the interface. Assuming the standard values of the gold permittivity as $\varepsilon_m$ = - 10.2 at $\lambda$ = 610 nm and $\varepsilon_m$ = -8.6 at $\lambda$ = 590 nm [37], and using the permittivity of the dielectric film, $\varepsilon_d$, as the fitting parameter, the condition for the excitation / decoupling of a -1$^{st}$ order SPP is

$$sin\theta = \frac{\lambda}{d} - \sqrt{\frac{\epsilon_m \epsilon_d}{\epsilon_m + \epsilon_d}}.$$

At $\varepsilon_d$ = 1.52, estimations yield: $\theta$ = 24° at λ = 610 nm and $d$= 350 nm; $\theta$ = 19° at λ = 590 nm and $d$= 350 nm; and $\theta$ = 6.6° λ = 610 nm, $d$= 420 nm which are reasonably close to the experimental observations of Figs. 3 (a, b) and (d). This confirms that we observe the efficient near-field emission from both electric and magnetic emitters to SPPs which are decoupled due to the periodic conditions.

The results obtained from the flat structure S2 are shown in Figure 4. Since in the fluorometer setup the readings are extremely sensitive to the prism position, we carefully adjust it to achieve the highest emission at various orientation angles, and record both s and p polarizations. Typical spectra recorded at the angle close to the SPP decoupling angle are shown in Fig. 4(b). While the s-polarized emission is very weak at all orientations tested, the p polarization is significant in a relatively narrow range of orientation angles. Such a maximum in the p –polarization intensity is observed for both electric and magnetic dipole emission, confirming coupling of both types of emitters with SPPs in flat films.

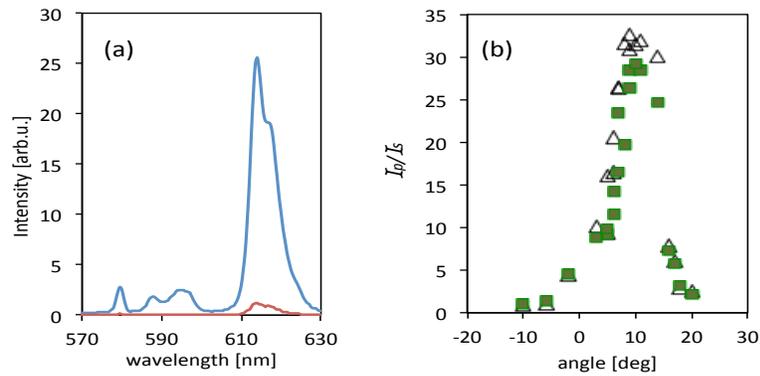

Fig. 4. Experiment with the flat film on the prism. (a) Spontaneous emission spectra recorded at p (blue) and s (red) polarization in the range of the SPP decoupling. (b) The ratio $I_p / I_s$ measured at the maximum of the electric transition, 613 nm (triangles) and the maximum of the magnetic transition 593 nm (squares).

In conclusion, we demonstrated that both electric and magnetic dipoles in the vicinity of plasmonic films can be efficiently coupled with SPPs, which results in an enhanced emission observed at the SPP decoupling angles. While for a particular system of $Eu^{3+}$ this approach is not expected to significantly affect the branching magnetic/electric ratio (since both transitions are from the same level), it can provide opportunities to enhance radiative transitions *vs* non-radiative decay in various systems.